%Paper: hep-lat/9304015
%From: Simon Catterall <S.M.Catterall@amtp.cam.ac.uk>
%Date: Tue, 27 Apr 93 10:31:50 +0100

%% jnl macro follows

%%                      	JNL.TEX
%%
%%                This is JNL.TEX Version 0.3 as of 6/12/85.
%%
%%	This is a set of TeX 82 macros designed to produce scientific
%%	papers with a minimum of fuss and using as much of plain.tex as
%%	possible.  The user need only know what is in the TeXbook, and
%%	the macros under ``user definitions'' below.  Also, the user
%%	definitions are intended to be as simple as possible, so that
%%	the user may change them as desired.

%%
%%  Font definitions suitable for the IMAGEN (Written by Tony Kennedy)
%%
 
%  Define a whole menagerie of pseudo-12pt fonts
 
\font\twelverm=cmr10 scaled 1200    \font\twelvei=cmmi10 scaled 1200
\font\twelvesy=cmsy10 scaled 1200   \font\twelveex=cmex10 scaled 1200
\font\twelvebf=cmbx10 scaled 1200   \font\twelvesl=cmsl10 scaled 1200
\font\twelvett=cmtt10 scaled 1200   \font\twelveit=cmti10 scaled 1200
 
\skewchar\twelvei='177   \skewchar\twelvesy='60
 
%  Define \...point macros to change fonts and spacings consistently
 
\def\twelvepoint{\normalbaselineskip=12.4pt
  \abovedisplayskip 12.4pt plus 3pt minus 9pt
  \belowdisplayskip 12.4pt plus 3pt minus 9pt
  \abovedisplayshortskip 0pt plus 3pt
  \belowdisplayshortskip 7.2pt plus 3pt minus 4pt
  \smallskipamount=3.6pt plus1.2pt minus1.2pt
  \medskipamount=7.2pt plus2.4pt minus2.4pt
  \bigskipamount=14.4pt plus4.8pt minus4.8pt
  \def\rm{\fam0\twelverm}          \def\it{\fam\itfam\twelveit}%
  \def\sl{\fam\slfam\twelvesl}     \def\bf{\fam\bffam\twelvebf}%
  \def\mit{\fam 1}                 \def\cal{\fam 2}%
  \def\tt{\twelvett}
  \textfont0=\twelverm   \scriptfont0=\tenrm   \scriptscriptfont0=\sevenrm
  \textfont1=\twelvei    \scriptfont1=\teni    \scriptscriptfont1=\seveni
  \textfont2=\twelvesy   \scriptfont2=\tensy   \scriptscriptfont2=\sevensy
  \textfont3=\twelveex   \scriptfont3=\twelveex  \scriptscriptfont3=\twelveex
  \textfont\itfam=\twelveit
  \textfont\slfam=\twelvesl
  \textfont\bffam=\twelvebf \scriptfont\bffam=\tenbf
  \scriptscriptfont\bffam=\sevenbf
  \normalbaselines\rm}
 
%	tenpoint

%%
%%	Various internal macros
%%
 
\def\beginlinemode{\endmode
  \begingroup\parskip=0pt \obeylines\def\\{\par}\def\endmode{\par\endgroup}}
\def\beginparmode{\endmode
  \begingroup \def\endmode{\par\endgroup}}
\let\endmode=\par
{\obeylines\gdef\
{}}
\def\singlespace{\baselineskip=\normalbaselineskip}

\def\oneandahalfspace{\baselineskip=\normalbaselineskip
  \multiply\baselineskip by 3 \divide\baselineskip by 2}
\def\doublespace{\baselineskip=\normalbaselineskip \multiply\baselineskip by 2}

\newcount\firstpageno
\firstpageno=2
\footline={\ifnum\pageno<\firstpageno{\hfil}\else{\hfil\twelverm\folio\hfil}\fi}
\let\rawfootnote=\footnote		% We must set the footnote style
\def\footnote#1#2{{\rm\singlespace\parindent=0pt\rawfootnote{#1}{#2}}}
\def\raggedcenter{\leftskip=4em plus 12em \rightskip=\leftskip
  \parindent=0pt \parfillskip=0pt \spaceskip=.3333em \xspaceskip=.5em
  \pretolerance=9999 \tolerance=9999
  \hyphenpenalty=9999 \exhyphenpenalty=9999 }
\def\dateline{\rightline{\ifcase\month\or
  January\or February\or March\or April\or May\or June\or
  July\or August\or September\or October\or November\or December\fi
  \space\number\year}}
\def\received{\vskip 3pt plus 0.2fill
 \centerline{\sl (Received\space\ifcase\month\or
  January\or February\or March\or April\or May\or June\or
  July\or August\or September\or October\or November\or December\fi
  \qquad, \number\year)}}
 
%%
%%	Page layout, margins, font and spacing (feel free to change)
%%

\parskip=\medskipamount
\twelvepoint		% selects twelvepoint fonts (cf. \tenpoint)
\doublespace		% selects double spacing for main part of paper (cf.
			%	\singlespace, \oneandahalfspace)
\overfullrule=0pt	% delete the nasty little black boxes for overfull box
 
%%
%%	The user definitions for major parts of a paper (feel free to change)
%%
 
\def\preprintno#1{
 \rightline{\rm #1}}	% Preprint number at upper right of title page
 
\def\title			%  Title on title page
  {\null\vskip 3pt plus 0.2fill
   \beginlinemode \doublespace \raggedcenter \bf}
 
\def\author			%  Author(s) name(s)  on title page
  {\vskip 3pt plus 0.2fill \beginlinemode
   \singlespace \raggedcenter}
 
\def\affil			% Affiliations (can intermix with \author)
  {\vskip 3pt plus 0.1fill \beginlinemode
   \oneandahalfspace \raggedcenter \sl}
 
\def\abstract			% Begin abstract
  {\vskip 3pt plus 0.3fill \beginparmode
   \doublespace \narrower ABSTRACT: }
 
\def\endtitlepage		% End title page, begin body of paper
  {\endpage			% 	This subsumes \body
   \body}
 
\def\body			% Begin text body;  can be used to end
  {\beginparmode}		% \title, \author, \affil, \abstract,
				% \reference, or \figurecaption modes
 
\def\head#1{			% Head;  NOTE enclose the text in {}
  \filbreak\vskip 0.5truein	%  e.g., \head{I. Introduction}
  {\immediate\write16{#1}
   \raggedcenter \uppercase{#1}\par}
   \nobreak\vskip 0.25truein\nobreak}
 
\def\subhead#1{			% Subhead;  NOTE enclose the text in {}
  \vskip 0.25truein		% e.g., \subhead{A. History of the Problem}
  {\raggedcenter #1 \par}
   \nobreak\vskip 0.25truein\nobreak}
 
\def\refto#1{$^{#1}$}		% For references in text as superscript
 
\def\references			% Begin references -- basic format is Phys Rev
  {\head{References}		% I.e., volume, page, year (space after commas).
   \beginparmode
   \frenchspacing \parindent=0pt \leftskip=1truecm
   \parskip=8pt plus 3pt \everypar{\hangindent=\parindent}}

\gdef\refis#1{\indent\hbox to 0pt{\hss#1.~}}	% Ref list numbers.
 
\gdef\journal#1, #2, #3, 1#4#5#6{		% Journal reference.  Comma sets
    {\sl #1~}{\bf #2}, #3 (1#4#5#6)}		% off: name, vol, page, year
 
\def\refstylenp{		% Nucl Phys(or Phys Lett) ref style: V, Y, P
  \gdef\refto##1{ [##1]}				% Reference in text []
  \gdef\refis##1{\indent\hbox to 0pt{\hss##1)~}}	% Ref list numbers)
  \gdef\journal##1, ##2, ##3, ##4 {			% Journal reference
     {\sl ##1~}{\bf ##2~}(##3) ##4 }}
 
\def\refstyleprnp{		% Input like pr, output like np!!
  \gdef\refto##1{ [##1]}				% Reference in text []
  \gdef\refis##1{\indent\hbox to 0pt{\hss##1)~}}	% Ref list numbers)
  \gdef\journal##1, ##2, ##3, 1##4##5##6{		% Journal reference
    {\sl ##1~}{\bf ##2~}(1##4##5##6) ##3}}

\def\np{\journal Nucl. Phys., }

\def\endreferences{\body}
 
\def\figurecaptions		% Begin figure captions
  {\endpage
   \beginparmode
   \head{Figure Captions}
}
 
\def\endfigurecaptions{\body}
 
\def\endpage			%  Eject a page
  {\vfill\eject}
 
\def\endpaper			%  Ways to say goodbye
  {\endmode\vfill\supereject}

\def\endit
  {\endpaper\end}

%%
%%	Various little user definitions
%%
 
\def\ref#1{Ref. #1}			% 	for inline references
			% 	ditto

		% For citation of equation numbers
	%	ditto
			%	ditto
			%	ditto
			%	ditto
			%	ditto
\def\frac#1#2{{\textstyle #1 \over \textstyle #2}}

\def\sla{\raise.15ex\hbox{$/$}\kern-.57em}
\def\leaderfill{\leaders\hbox to 1em{\hss.\hss}\hfill}
\def\twiddle{\lower.9ex\rlap{$\kern-.1em\scriptstyle\sim$}}
\def\bigtwiddle{\lower1.ex\rlap{$\sim$}}
\def\gtwid{\mathrel{\raise.3ex\hbox{$>$\kern-.75em\lower1ex\hbox{$\sim$}}}}
\def\ltwid{\mathrel{\raise.3ex\hbox{$<$\kern-.75em\lower1ex\hbox{$\sim$}}}}
\def\square{\kern1pt\vbox{\hrule height 1.2pt\hbox{\vrule width 1.2pt\hskip 3pt
   \vbox{\vskip 6pt}\hskip 3pt\vrule width 0.6pt}\hrule height 0.6pt}\kern1pt}

%%%%%%%%%%%%%%%%%%%%%%%%%%%%%%%%%%%%%%%%%%%%%%%%%%%%%%%%%%%%%%%%%%%

%% postscript files appended at end of text in tar compressed uuencoded form

\doublespace

\preprintno{ILL--(TH)--93--5}
\preprintno{DAMTP--93--16}

\title{The XY model on a dynamical random lattice}
 
\author{Simon M. Catterall}

\affil
Department of Applied Maths and Theoretical Physics
University of Cambridge
Silver St. Cambridge CB3 9EW
England

\author{John B. Kogut}

\affil
Loomis Laboratory of Physics
University of Illinois at Urbana--Champaign
1110 West Green Street
Urbana, Illinois 61801

\author{Ray L. Renken}
 
\affil
Department of Physics
University of Central Florida
Orlando, Florida 32816
 
\abstract
{We study the XY model on a lattice with fluctuating connectivity.
The expectation is that at an appropriate critical point such a system
corresponds to a compactified boson coupled to 2d quantum gravity.
Our simulations focus, in particular, on the important topological 
features of the system. The results lend strong support to the
two phase structure predicted on the basis of analytical
calculations.
A careful finite size scaling analysis yields estimates for
the scaling exponents in the low temperature phase.}

\vfill
\eject

\subhead{Introduction}

In recent years considerable progress has been made in understanding
the nature of some simple systems encorporating two dimensional quantum
gravity. These systems are thought to correspond to
string theories out of the critical dimension. 
Two approaches have been followed which are at first sight quite
different. The first of these employs techniques 
borrowed from conformal field theory 
to calculate the spectrum of anomalous dimensions in several
simple models [1-2]. In the second method a regularisation of the
continuum functional integrals is made in terms of random triangulations [3-5].
Where it has been possible to solve the models analytically the two
approaches have been in complete agreement. Furthermore,
the triangulated models may be studied nonperturbatively by
numerical simulation. Whilst the analytical work gives only
information on the scaling behaviour of matter field expectation
values, the numerical studies allow us to probe the geometrical
and fractal structure of the worldsheet. The lattice models may
also be used for matter field central charges $c$ greater than unity where
the continuum methods break down.

The marginal case when $c=1$ can be realised by
employing a single scalar field. If the latter model is
compactified by imposing that the field be periodic in some
interval, no exact solution
has been found. However there are strong theoretical arguments to
suggest that the model may exist in two phases; one in which the scalar
field is non-propagating and is described by pure ($c=0$) gravity and
another corresponding to a massless, propagating scalar field with
unit central charge [6].

The lattice model is also interesting from a statistical mechanics
point of view. The model in flat space (regular lattice) is known to
undergo a Kosterlitz-Thouless KT phase transition at some critical coupling
to a massless low temperature phase with a set of continuously varying
critical exponents. This transition is driven by the condensation
of topological defects (vortices) corresponding to 
field configurations with non-zero winding number.
It is interesting to ask how this picture is modified by the
encorporation of lattice (or metric) fluctuations. Naively, one might
imagine that performing an annealed sum over lattices might
strongly influence the effective forces between vortices and
hence alter the character of any phase transition. Much of the
results we present later will be concerned with elucidating the role
such topological configurations play on a dynamical lattice.

\subhead{Discussion of the regular lattice model}

Consider initially the model on a fixed, regular lattice. The latter
is taken to be the dual to a regular triangulation -- a $\phi^3$-graph
of toroidal topology and possessing $N$ nodes.
The partition function defining the model is
$$Z=\sum_{\theta_i}e^{-S\left(\beta\right)}\eqno(1)$$
with the lattice action $S\left(\beta\right)$ given by
$$S\left(\beta\right)=-\beta\sum_{\left\langle ij\right\rangle}\cos\left(
\theta_i-\theta_j\right)\qquad \theta_i\in 0\cdots 2\pi\eqno(2)$$
Using techniques drawn from conformal field theory it is possible to
show that the critical region of this theory can be described in terms of
two sets of operators with anomalous dimensions $\{\Delta_n,\overline{
\Delta}_m\}$ (see for example [7]). 
$$\Delta_n={1\over 2}\left(1\over 2\pi\beta_{\rm eff}\right)n^2\eqno(3)$$
$$\overline{\Delta}_m={1\over 2}\left(2\pi\beta_{\rm eff}\right)m^2\eqno(4)$$
The former operators ($\Delta_n$) correspond to spin waves, whilst the
latter ($\overline{\Delta}_m$) are to be identified with vortices.
The effective coupling constant $\beta_{\rm eff}$ is determined by
details on the scale of the lattice cutoff, for example the lattice type 
and bare
lattice coupling $\beta$. However, for large coupling $\beta$ we expect 
$$\beta_{\rm eff}\sim\beta ,\qquad\beta\to\infty\eqno(5)$$
The vortex operators are seen to be irrelevant operators at large 
$\beta_{\rm eff}\sim\beta$ and the critical theory can be understood
in terms of its spin wave content. In this region we expect that 
the theory is massless and the two point
function corresponding to spin wave excitations behaves as
$$\left\langle e^{i\theta\left(0\right)}e^{-i\theta\left(r\right)}\right\rangle
\sim r^{-2\Delta_1}\sim r^{-{1\over 2\pi\beta_{\rm eff}}}\eqno(6)$$
However for small $\beta$ this situation is reversed and
the vortex operators are relevant and dominate
the long distance physics. A phase transition is thus expected
between a region where spin waves essentially exhaust the
physically relevant degrees of freedom and one where the ground
state is populated by vortices.
This transition will occur when the
the anomalous dimension of the lowest vortex operator $\overline{\Delta}_1$
is just marginal ($\overline{\Delta}_1=2$). This
yields an estimate for the critical coupling
$$\beta_{\rm eff}^c={2\over \pi}\eqno(7)$$
A careful renormalisation group analysis [8] shows that
the phase transition at $\beta_c$ is of an unusual type -- the free energy
and all its derivatives are continuous at $\beta_c$ whilst the
correlation length and susceptibility possess essential
singularities as the transition is approached from the disordered phase. 
$$\xi\sim ae^{{b\over \left(\beta_c-\beta\right)^\nu}}\eqno(8)$$
$$\chi\sim ce^{{d\over \left(\beta_c-\beta\right)^\nu}}\eqno(9)$$
The exponent $\nu$ takes the value $\nu=1/2$.
Furthermore, the dimension of the lowest spinwave operator is then
$$\Delta_1={1\over 8}, \qquad \beta=\beta_c\eqno(10)$$

The presence of such a phase transition can be inferred in
another way by considering the free energy of such a vortex configuration.
At large $\beta$ the cosine action may be replaced by an equivalent
gaussian action, provided we retain the periodicity of the $\theta$
coordinates. A simple vortex configuration would be
$\theta=m\phi$ in an $\left(r,\phi\right)$ polar coordinate system. 
This yields an action for the vortex
of 
$$S_{\rm vortex}=-{\beta_{\rm eff}}\pi n^2\ln\left(L\over a\right)\eqno(11)$$
where $L$ is the infrared and $a$ the ultraviolet cutoff. Since there
is a choice of $\left(L/a\right)^2$ positions on a cubic lattice
at which to place the centre of such a vortex (with $n=1$) the contribution to
the free energy of such a configuration is
$$\delta F=\left(\pi\beta_{\rm eff}-2\right)\ln\left(L\over a\right)\eqno(12)$$
Clearly for large $\beta_{\rm eff}$ vortices will be
irrelevant, whilst for small coupling they will dominate,
disordering the ground state and forcing a finite correlation
length. 

The presence of such non-trivial field configurations is signaled
by a non-zero vorticity $V$ defined in the continuum by
$$V={1\over 2\pi}\int_C \nabla\theta .ds \eqno(13)$$
We have employed the naive transcription of this object onto the
lattice
$$V=\sum_{\rm loops}^{N_L}\sum_j\left(\theta_{j+1}-\theta_j\right)\bmod {2\pi}
\eqno(14)$$
The outer sum is carried out over all the $N_L$ loops of the $\phi^3$-graph, 
whilst
the inner sum corresponds to computing the net change in angle
on traversing a given loop.
In practice, the total vorticity vanishes as a consequence of the
boundary conditions, so we measure the modulus of the local 
vorticity in our simulations.
Clearly, at large $\beta$, where all the spins are locally
correlated, this is likely to be a good approximation to 
the continuum expression, but one might
worry about the situation at small coupling. This is, of course,
the usual problem of defining topology on a lattice. The problem
is to construct lattice observables which go smoothly over into
continuum topological quantities. Typically, the lattice operators
receive large contributions from fluctuations at order the cutoff
which have no place in the continuum expectation values.
One of the standard methods to measure
topological charge for lattice QCD is by a so-called cooling method [9],
whereby the action is locally minimised to remove ultraviolet
fluctuations. We have tested such a method in this situation and found
that for a wide range of couplings close to criticality the
winding number remains constant under cooling giving us confidence that we are
not seeing lattice artefacts.

\subhead{Fluctuating lattice model}

The simulations described later will concern themselves with
a nonperturbative study of this model when formulated
on a dynamical random lattice. As we have described this
prescription is equivalent (at least when $c<1$) to
coupling the spin model to gravity.
Since the XY model has $c=1$ at large $\beta$, the
KPZ framework [1] marginally applies and we can attempt to
derive some
predictions for this model in the presence of such metric 
fluctuations.
Specifically, if we take some operator $O\left(x\right)$ and integrate it
over the surface (to get a reparametrisation invariant object), the
expectation value will scale with surface area as
$$\left\langle \int d^2x\sqrt{g}O\left(x\right)\right\rangle_{\rm gravity}
\sim A^{1-D}, \qquad A\to\infty\eqno(15)$$
The KPZ formula then relates this gravitational scaling
dimension $D$ to the anomalous dimension
$\Delta$ of $O\left(x\right)$. 
$$D=\sqrt{\Delta/2},\qquad c=1\eqno(16)$$
Similarly, an integrated correlation function or susceptibility
conjugate to $O\left(x\right)$ would scale like
$$\left\langle\int d^2x\sqrt{g}O\left(x\right)
\int d^2x^\prime\sqrt{g^\prime}
O\left(x^\prime\right)\right\rangle_{\rm gravity}\sim A^{\left(2-2D\right)}
\eqno(17)$$
To adopt a conventional normalisation of the susceptibility we would
divide by a power of the area to get
$$\chi\left(O\right)\sim 
A^{\left(1-2D\right)}=A^{\left(1-\sqrt{2\Delta}\right)}\eqno(18)$$
which is to be compared with the usual finite size scaling 
in the absence of gravitational fluctuations
$$\chi^{\rm flat}\left(O\right)\sim A^{\left(1-\Delta\right)}\eqno(19)$$

For the gravity coupled system the condition for 
relevance of an operator $O$ is now that its gravitational
scaling dimension $D\left(O\right)=1$, which in the case of
the vortex operators $O=\overline{\Delta}_1$ yields
the condition that $\overline{\Delta}_1=2$. This
is precisely the same condition that determined the critical
coupling in flat space. Thus this continuum analysis
would predict that the gravitational dressing of the vortex operators
does not change the position of the phase transition. However it is
not clear that the transition is still of KT type. We will address
this question again when interpreting our results.

The simulation of this dynamical model is effected by means of
two forms of update. To simulate the sum over spin configurations
we have employed the cluster update pioneered by Wolff [10].
This proves to be a very efficient method for quasi-massless
theories like the XY model.
The sum over graphs is effected by a local procedure detailed in
a previous paper [11]. We have simulated lattice volumes
from $N=100$ to $N=5000$, typically using $O\left(10^5\right)$
sweeps of the lattice per coupling $\beta$.
Errors were assessed by the usual rebinning procedure.

\subhead{Results and analysis}

Since the continuum arguments suggest that the transition on the dynamical
lattice may be rather similar to the regular (flat space) theory, we chose
to do some moderate simulations of the latter to compare with
the dynamical results. Fig.1 shows a plot of the specific heat $C$
$$C={1\over N}\left(\left\langle S^2\right\rangle -\left\langle S\right\rangle
^2\right)\eqno(20)$$
as a function of lattice coupling $\beta$. The plot also shows the
topological susceptibility $\chi_v$ determined by
$$\chi_v={1\over N_L}\left(\left\langle V^2\right\rangle -\left\langle V
\right\rangle^2\right)\eqno(21)$$
Clearly, both show peaks in the neighbourhood of $\beta_c\sim 1.6$ which
may be taken as a signal for some form of critical behaviour. However, the
specific heat rapidly saturates with increasing system volume and hence
is not a good observable for determining the nature of any phase
transition.
Furthermore, in the case of the dynamical system the peak in $C$ occurs
at a substantially different coupling than our best estimate
of $\beta_c$. Similarly, the spin
susceptibility increases monotonically with $\beta$ and yields no
information on the position of any phase transition. 

In contrast, the topological susceptibility being
determined by the degrees of freedom which are believed to drive
the phase transition, is a much more reliable indicator of
criticality in the system. Indeed, we have taken the peak in $\chi_v$
as our best estimate of the critical point. From the largest
regular lattices we have considered we estimate the critical lattice
coupling to be $\beta_c=1.7(1)$.
As we detail below this
produces estimates for the critical exponents on a regular lattice
which are in agreement with analytical calculations.

Fig 2. shows the vorticity $V$ for a 3000 node dynamical lattice over a range
of couplings spanning the critical region. The absence of
any discontinuities makes it unlikely that there are any first
order transitions in the system. However the topological susceptibility
is much more interesting and is shown in fig. 3 for the full
range of system sizes we studied. The first point to note is
that the critical coupling, as dictated by the peak in $\chi_v$, has
substantially shifted from its value on the regular (flat space) lattice 
$\beta_c=1.7$ to a value which we estimate from the peak on the largest lattice
to be $\beta_c=2.3(1)$. Note that this is in agreement with an earlier study
which utilised an entirely different method to derive a value for $\beta_c$
[12]. 

However, at first sight, this renormalisation of the
coupling would appear to be at variance with
the continuum calculations which indicate that transition coupling is
not changed by the metric fluctuations.
In an effort to cast some light on this we simulated an XY model on a single
quenched lattice deriving from a simulation
of pure gravity. In this case the peak in $\chi_v$ again occurred in the
vicinity of $\beta_c=2.3$. We interpret this situation
in the following way.
The transition occurs at some value of $\beta_{\rm eff}$ which is
identical on both flat and fluctuating geometries. However the corresponding
bare lattice coupling may be different
in the regular and dynamical lattice cases (since their local
lattice structure is radically different). Furthermore, it is then conceivable 
that
the KPZ predictions (eqn. 18) for the fluctuating lattice will still hold
true at the appropriate critical point (and
indeed our later results will lend support to this conclusion).

The second observation to be made from fig. 3 is 
that $\chi_v$ shows strong finite volume behaviour for large $\beta$. This
is also true of the model restricted to a regular lattice and comes about
for the following reason.
Consider the regular lattice topological susceptibility $\chi_v^{\rm flat}$ 
as a (euclidean) time $t$
integral over a vortex-vortex correlator.
$$\chi_v^{\rm flat}=\sum_t\sum_s \bigl\vert\left\langle v_0|
s\right\rangle\bigr\vert^2
e^{-E_s t}\eqno(22)$$
The states $s$ are eigenstates of the lattice Hamiltonian with energies $E_s$. 
Clearly
the integral will be dominated by the lightest state $s_0$ which
couples to the local vortex operator $v_0$. It will be
this state which will govern the finite size scaling of the
susceptibility. If we regard the matrix elements 
$\left\langle v_0|s\right\rangle$
now as expectation values in a path integral, we can use general symmetry
principles to determine which of them, in principle, are nonvanishing.
Consider the overlap of a spinwave state $s=e^{i\theta}$ on the local
vorticity $v_0$. If we perform a parity transformation of the form
$$x_1\to -x_1,\qquad x_2\to x_2,\qquad \theta\to\theta\eqno(23)$$
the vorticity changes sign but the spin wave state is invariant. Thus, provided
parity is not spontaneously broken,
the spinwave and vortex sectors are completely orthogonal. However if
we take as operator the modulus of the local vorticity (which is
the case in our numerical simulations) then both operators are
invariant under this parity transformation  and hence we would
expect on these general grounds that the lightest state coupling to
such an operator would again be the simple spin wave state. Thus the finite
size scaling of the topological susceptibility will be governed by the
same critical exponent $\Delta_1$ as for the spin susceptibility.
Similarly in the gravitationally dressed case of the dynamical
ensemble the two susceptibilities will scale with the same exponent.
A measurement of both quantities gives us independent estimates of this 
scaling exponent.

This picture is borne out by our numerical data. Fig. 4 reveals an
analysis of the finite volume behaviour of the spin 
susceptibility $\chi_s$ for both regular and fluctuating lattices.
As we emphasised, we expect 
that its finite size
scaling will be dominated by the lightest spin wave state ($n=1$).

The regular lattice data shown derive from runs at the transition
point $\beta_c=1.7$
We fit our data from lattice sizes $N=968$ to $N=5000$ to the form
$$\chi\sim a+bN^\omega\eqno(24)$$
using a simple nonlinear least squares procedure. The fit yields
$\omega^{\rm flat}=0.855(3)$ with $\chi^2/{\rm dof}=0.4$ with 55\% confidence.
This number is to be contrasted with the prediction $\omega^{\rm flat}=0.875$
dictated by the usual KT and CFT arguments.
However the fluctuating lattice gives the fit $\omega=0.53(13)$ with
$\chi^2/{\rm dof}=0.04$ at 85\% confidence. This, in turn, lies
remarkably close to the KPZ prediction of $\omega=0.5$ derived from
eqn. 18 with $\Delta_1=1/8$.

Similar conclusions can be drawn in the case of the topological
susceptibility $\chi_v$ (fig. 5). Here, the regular lattice fit
gives $\omega^{\rm flat}=0.843(31)$, $\chi^2/{\rm dof}=0.02$ with
a confidence level of 89\%, which is consistent with the
spin susceptibility measurement and compatible with the theoretical
prediction. The dynamical lattice yields $\omega=0.64(15)$, $\chi^2/{\rm dof}=
0.1$ at 79\% confidence.  This again is in reasonable agreement with
the KPZ number of $0.5$ and is inconsistent (at two standard deviations)
with the flat space scaling exponent $\omega=0.875$.

Clearly then, we have seen very good agreement with the usual KT assignments
in the case of flat space. This gives us confidence that our method for
locating the transition coupling is a good one. Furthermore, the
dynamical exponents are dramatically different and lie 
intriguingly close to the naive KPZ prediction
of the continuum theory, even the critical lattice coupling for
the dynamical ensemble is significantly different from its flat space value.

As an independent check we have examined the spin correlator $g\left(r\right)$
for the regular lattice at our
best estimate for $\beta_c=1.7$. Fig.6 shows an
`effective' $\eta$-plot determined from the
spin correlator on a lattice with $N=2888$ by assuming
power law behaviour 
$$\eta\left(r\right)={\ln\left({g\left(r+1\right)\over g\left(r\right)}\right)
\over \ln\left({r+1\over r}\right)}\eqno(25)$$
We expect finite size effects to be important when $r>15$ and lattice spacing
errors to be present for small $r$, but the figure clearly shows a rather
stable plateau for $\eta$ within this region. Indeed the plot of fig. 7
allows a fit for $\eta$ in this range yielding $\eta=0.247(1)$ with
$\chi^2/{\rm dof}=1.4$. This is to be compared with the
predition of $\eta=2\Delta_1=0.25$ from KT theory.
The `effective' mass computed in the analogous way on the assumption of 
an exponential behaviour for $g\left(r\right)$ is shown also in fig. 6.
Clearly there is no stable region for fitting here and we interpret
this as strong indication that at the critical point of the regular
lattice XY model there are power law correlations.

Contrast these conclusions with fig. 8 which shows the same quantities
computed with the ensemble of random graphs. In this case we have
to be more careful in defining the correlation function $g\left(r\right)$.
On the lattice we define $g$ as follows
$$g\left(r\right)=\left\langle{1\over n\left(r\right)}\sum_{ij}\sigma_i\sigma_j
\delta\left(d_{ij}-r\right)\right\rangle\eqno(26)$$
The function $d_{ij}$ is just the geodesic distance (minimal length walk)
on a given graph between sites $i$ and $j$, whilst
$$n\left(r\right)=\sum_{ij}\delta\left(d_{ij}-r\right)\eqno(27)$$
We average over spin configurations $\sigma_i=e^{i\theta_i}$ and graphs $G_N$.
For a fixed lattice this just reduces to the usual definition of the
propagator.
Notice that the continuum version of this quantity is just
$$g\left(r\right)=\left\langle{1\over n\left(r\right)}\int d^2x 
\int d^2x^\prime 
\sqrt{g}\sqrt{g^\prime} 
\phi\left(x\right)\phi\left(x^\prime\right)\delta\left(d\left(g,x,x^\prime
\right)-r\right)\right\rangle_{{\phi}\rm +grav}\eqno(28)$$
This correlation function is explicitly reparametrisation invariant 
(if the fields $\phi$ are scalars) since the identity and length
of the geodesics $d\left(x,x^\prime,g\right)$ are reparametrisation
invariant. Therefore this correlation function is 
a bona fide observable for gravity coupled systems.
However, in this dynamical case, the effective power fit plot at
$\beta=2.3$ (fig. 8), is extremely
poor, and the exponential fit proves considerably better. Indeed, the
best fit we could produce (fig. 7) corresponded to a two 
exponential fit from distances
3 to 15 of the form
$$g\left(r\right)=a\exp{-br}+c\exp{-dr}\eqno(29)$$
with the results $a=0.23(1)$, $b=0.32(4)$, $c=0.69(2)$ and $d=0.048(2)$ with
a $\chi^2/{\rm dof}=0.28$ at 97\% confidence. The second
exponential presumably corresponds to an excited state, but at present it is
not at all clear why there is this qualitative difference in the
structure of this correlation function $g$ when the
system is coupled to gravity.

We have also attempted to fit the spin susceptibility for
the dynamical lattice simulations 
to the KT form of essential singularity. Using the data at $N=3000$ and
with $\beta_c=2.3$ as before, we have assumed $\nu=0.5$ and
attempted to fit the data in the disordered phase. Fig. 9 illustrates
this by plotting $\ln\chi_s$ as a function of $\left(\beta_c-\beta\right)^
{-0.5}$. The rounding visible in the plot corresponds to the onset of
finite size effects, but clearly the data set is not inconsistent
with a KT type singularity. However, it is perfectly possible to fit
the data reasonably well over the same range by a simple power fit and
objectively it is very difficult to assess which is the better.
This situation is well known in the case of the ordinary XY model
and it is only relatively recently that numerical simulations have
proven capable of resolving this problem [13]. With the lattice
sizes employed in this study, such a definitive conclusion is
impossible.
However we can say that a fit of the form
$$\chi\sim a\left(\beta_c-\beta\right)^{-\gamma}\eqno(30)$$
requires, typically, a very large coefficient $a\sim 40.0$ and power
$\gamma=3.5-4.0$. 

To examine the effective gravitational action induced by the coupling
of spins to the dynamical lattices, we also measured the fluctuation
in the local scalar curvature. This is just defined by
$$r^2\left(\beta\right)=\left\langle\sum_i^{N_L}{\left(6-l_i\right)^2\over 
l_i}\right\rangle\eqno(31)$$
where $l_i$ is the length of loop $i$ and the sum runs over all loops
in the $\phi^3$-graph.
Fig 10. shows this as a function of the bare lattice coupling for a variety
of lattice volumes. Clearly finite size effects are small, and the coupling
to matter enhances the production of singular geometries. Notice that the
the effects of this are almost constant throughout
the low temperature phase, perhaps suggesting that, as has been observed
before [14,15], the effective action for gravity may be governed 
only by the
central charge; the field content and anomalous dimensions seem
to have little influence on the gravitational sector (at least in
those models examined so far).
The figure also shows the same quantity for two Ising models confirming
this universality.

On the same figure we show a quantity $W$ corresponding to
the cross correlator of the local vorticity (again its modulus strictly)
with the ring length. 
$$W\left(\beta\right)={\left\langle\sum_i^{N_L}l_i|v_i|\right\rangle\over
\sum_i^{N_L}\left\langle l_i\right\rangle\left\langle |v_i|\right\rangle}-1
\eqno(32)$$

The signal is normalised by the disconnected part
to factor off the trivial $\beta$ dependence of the vorticity. Again a peak
is seen close to the initial peak in the
scalar curvature fluctuation $r^2$ and 
significantly to the left
of the estimated transition point. This would indicate a significant
tendency of vortices to bind to curvature defects on the lattice. The latter
can be understood qualitatively as a consequence of the logarithmic
divergence of the free energy of a single vortex. The free energy is lowered
if the vortex forms on a ring of large length (the ultraviolet cutoff a in
eqn. 11). This conclusion appears to be essentially unaltered when 
fluctuations are
taken into account.

Finally we have studied the clustering properties of vortices and
antivortices. To do this we introduce the joint probability distributions
$P_{vv}\left(r\right)$ and $P_{vav}\left(r\right)$. If we have
a vortex at the origin $r=0$, these measure the probabilities of finding
a vortex (in the case of $P_{vv}$) or antivortex ($P_{vav}$) at some
geodesic distance $r$ (measured with the respect to the loops). 
Fig. 11 illustrates this with histograms of these distributions in the
disordered phase and ordered phase. Clearly the vortex-vortex distribution
is essentially the same in both cases and corresponds to an approximately
uniform distribution of vortices over the plane. However a dramatic
difference is seen in the case of $P_{vav}$. In the disordered phase
the there is a no evidence of any real clustering, whilst in the
low temperature phase almost all antivortices are bound one lattice
spacing away. This is completely analogous to the situation in
flat space where the disordered phase corresponds to a uniform plasma
of vortices and antivortices, whilst in the low temperature phase
the vortices and antivortices exist only in bound pairs. The transition
is then pictured in this language as the dissociation of these dipole-like
pairs, whilst the peak in $\chi_v$ corresponds to a maximal
fluctuation between pairs as their effective binding decreases to zero.
Thus, the qualitative picture of a transition driven by condensation
of defects persists even when the defects are dressed by
gravitational fluctuations.

\subhead{Conclusions}

We have presented results for the XY model coupled to a fluctuating
lattice of fixed topology. These simulations are motivated by the need
to understand the coupling of 2d gravity to simple matter systems.
We have collected data both for the standard spin wave observables and also
the topological sector. The estimates of the critical exponents we
derive from finite size scaling analyses are in agreement with both
flat space KT predictions and the KPZ framework. 

We observe that the fluctuations in the geometry plateau in the low
temperature phase at a value commensurate with a two Ising system which
is consistent with the assignment of central charge one for this phase.
Furthermore we observe a strong binding of the vortices to regions 
of high curvature close to the critical point. The distribution functions
for vortex-vortex and vortex-antivortex pairs confirm the qualitative
picture of the two phases as given by KT theory in flat space. The
presence of metric fluctuations, whilst dressing operators, seems to
preserve the nature of the transition.

These results then confirm the phase structure argued for on the basis
of continuum calculations, and furnish yet further 
evidence of the validity of the dynamical triangulation prescription
(and its computer implementation)
in the marginal case of $c=1$ theories.

This work was supported, in part, by NSF grant PHY 92-00148 and some of the
numerical calculations were performed using the Florida State University
CRAY YMP. 
We acknowledge fruitful discussions with Ian Drummond and Sumit
Das.

\references

[1]  V. Knizhnik, A. Polyakov and A. Zamolodchikov, Mod. Phys. Lett.A3 819 1988.

[2]  J. Distler and H. Kawai, \np B321, 509, 1989.

[3]  J. Ambjorn, B. Durhuus and J. Frohlich, \np B257, 433, 1985.

[4]  F. David, \np B257, 543, 1985.

[5]  B. Boulatov, V. Kazakov, I. Kostov and A. Migdal, \np B275, 641, 1986.
     
[6]  D. Gross and I. Klebanov, \np B344, 475, 1990.

        D. Gross and I. Klebanov, \np B354, 475, 1990.

        D. Boulatov and V. Kazakov, Nucl. Phys. B (Proc. Suppl.) 25A (1992) 38.

[7]  Itzykson and Zuber, Statistical Field Theory II, CUP 1989.

[8]  J. Kosterlitz and D. Thouless, J. Phys. C6 (1973) 1181. 

        J. Kosterlitz, J. Phys. C7, (1974), 1046.

[9]  M. Teper, Phys. Lett. B202 (1988) 553.

[10]  U. Wolff, Phys. Rev. Lett. 62 (1989) 361.

[11]  S. Catterall, D. Eisenstein, J. Kogut and R. Renken, \np B366, 647, 1991.

[12]  C. Baillie and D. Johnston, COLO-HEP-286.

[13]  U. Wolff, \np B322, 759, 1989.

         R. Gupta and C. Baillie, Phys. Rev B45 (1992) 2883.

[14]  C. Baillie and D. Johnston, COLO-HEP-276.

      G. Thorleifsson, Nucl. Phys. B (Proc. Suppl.) 30 (1993) 787.

[15]  S. Catterall, R. Renken and J. Kogut, Phys. Lett. B292, (1992) 277.

\endreferences

\figurecaptions

[1]  Specific Heat $C$ and Topological Susceptibility $\chi_v$ for
     a regular lattice ($N=968$) as a function of bare coupling $\beta$.

[2]  Vorticity on a 3000 node dynamical lattice vs $\beta$.

[3]  Topological Susceptibility $\chi_v$ on a dynamical lattice.
     Lattice volumes are $N=100$ ($\times$), $N=1000$ ($\triangle$)
     $N=2000$ ($\diamond$), $N=3000$ ($\square$), $N=5000$ ($\circ$).

[4]  Spin Susceptibility $\chi_s$ for both regular ($\circ$) and 
     fluctuating ($\diamond$) lattices vs number of nodes $N$.
   
[5]  Topological Susceptibility $\chi_v$ for both regular ($\circ$) and
     fluctuating ($\diamond$) lattice vs number of nodes $N$.

[6]  Power ($\circ$) vs exponential ($\square$) fits for a regular lattice 
     $N=2888$ at $\beta=1.7$.

[7]  Spin Correlators for regular and fluctuating lattices at
     criticality together with power law and exponential fits respectively.

[8]  Power ($\circ$) vs exponential ($\square$) fits for a dynamical lattice
     $N=3000$ at $\beta=2.3$.

[9]  $\log{\left(\chi_s\right)}$ vs $t^{-1/2}$, $t=\beta_c-\beta$, together
     with KT linear fit.

[10] Fluctuation in local scalar curvature $r^2$ and cross correlator 
     of vorticity with loop length $W$ vs $\beta$ .

[11] Histograms of joint probability distributions for vortex-vortex
     ($P_{vv}\left(r\right)$) and vortex-antivortex ($P_{vav}\left(r\right)$)
     pairing for $\beta=3.0$ and $\beta=1.0$.

\endfigurecaptions

\endit

%% figures follow

#!/bin/csh -f
# Note: this uuencoded compressed tar file created by csh script  uufiles
# if you are on a unix machine this file will unpack itself:
# just strip off any mail header and call resulting file, e.g., figs.uu
# (uudecode will ignore these header lines and search for the begin line below)
# then say        csh figs.uu
# if you are not on a unix machine, you should explicitly execute the commands:
#    uudecode figs.uu;   uncompress figs.tar.Z;   tar -xvf figs.tar
#
uudecode $0
chmod 644 figs.tar.Z
zcat figs.tar.Z | tar -xvf -
rm $0 figs.tar.Z
exit
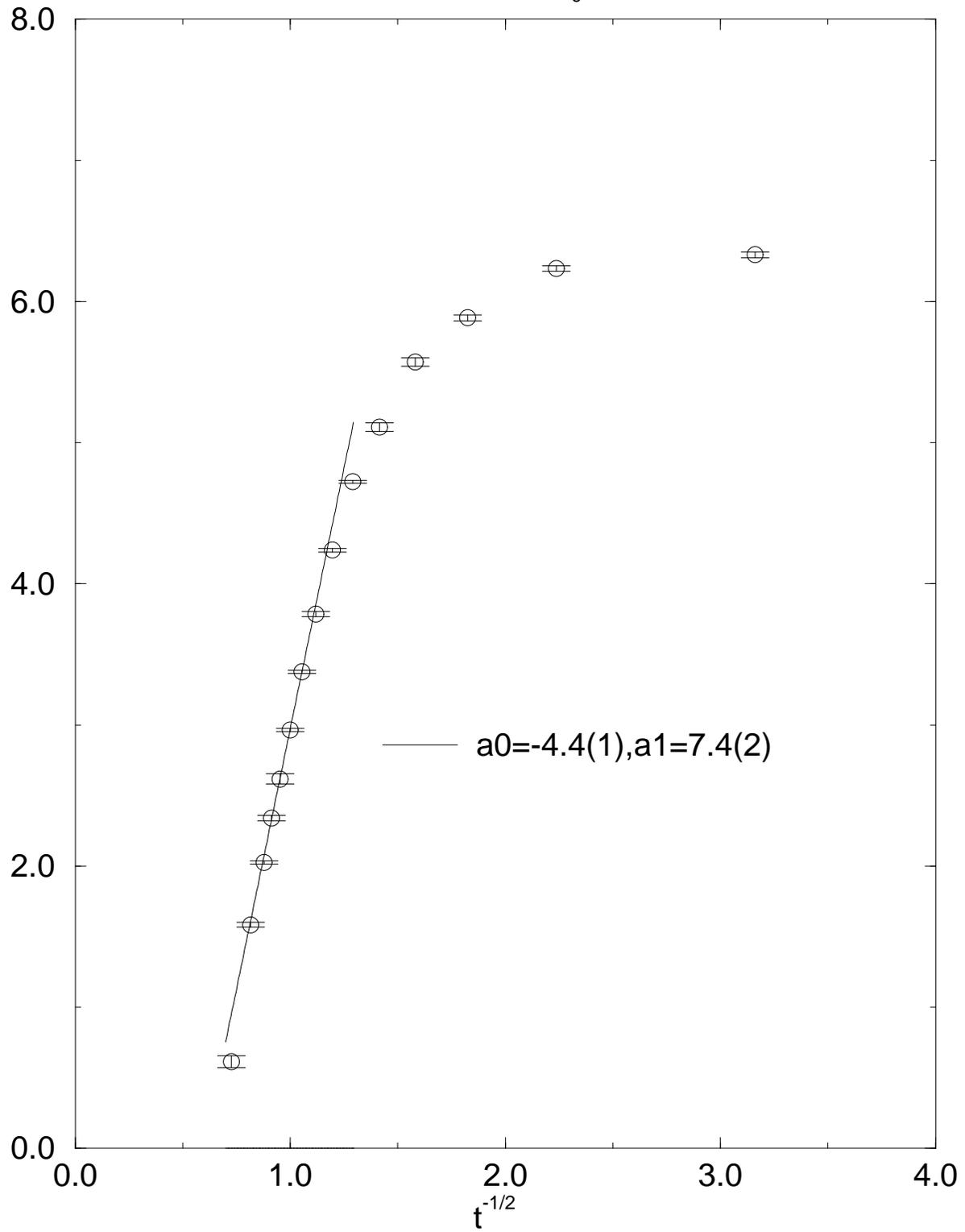
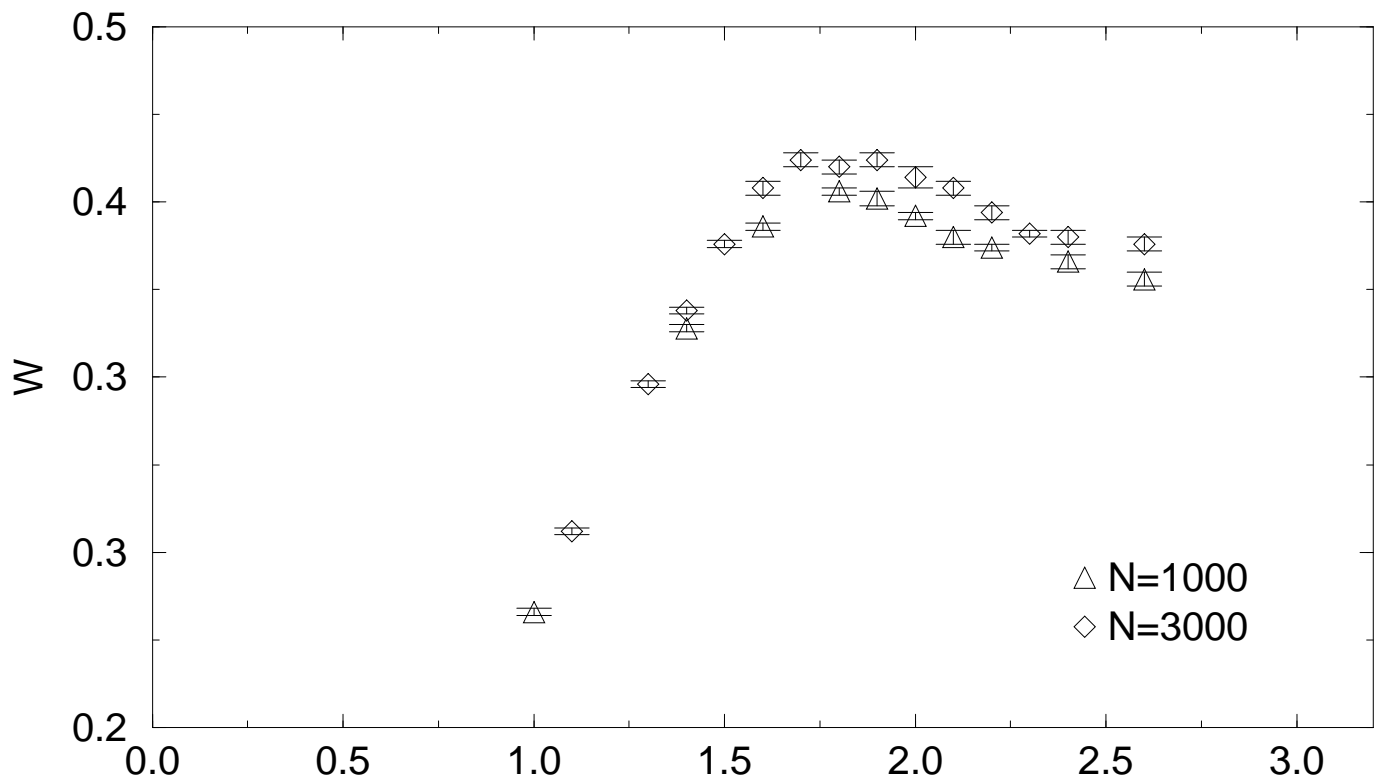
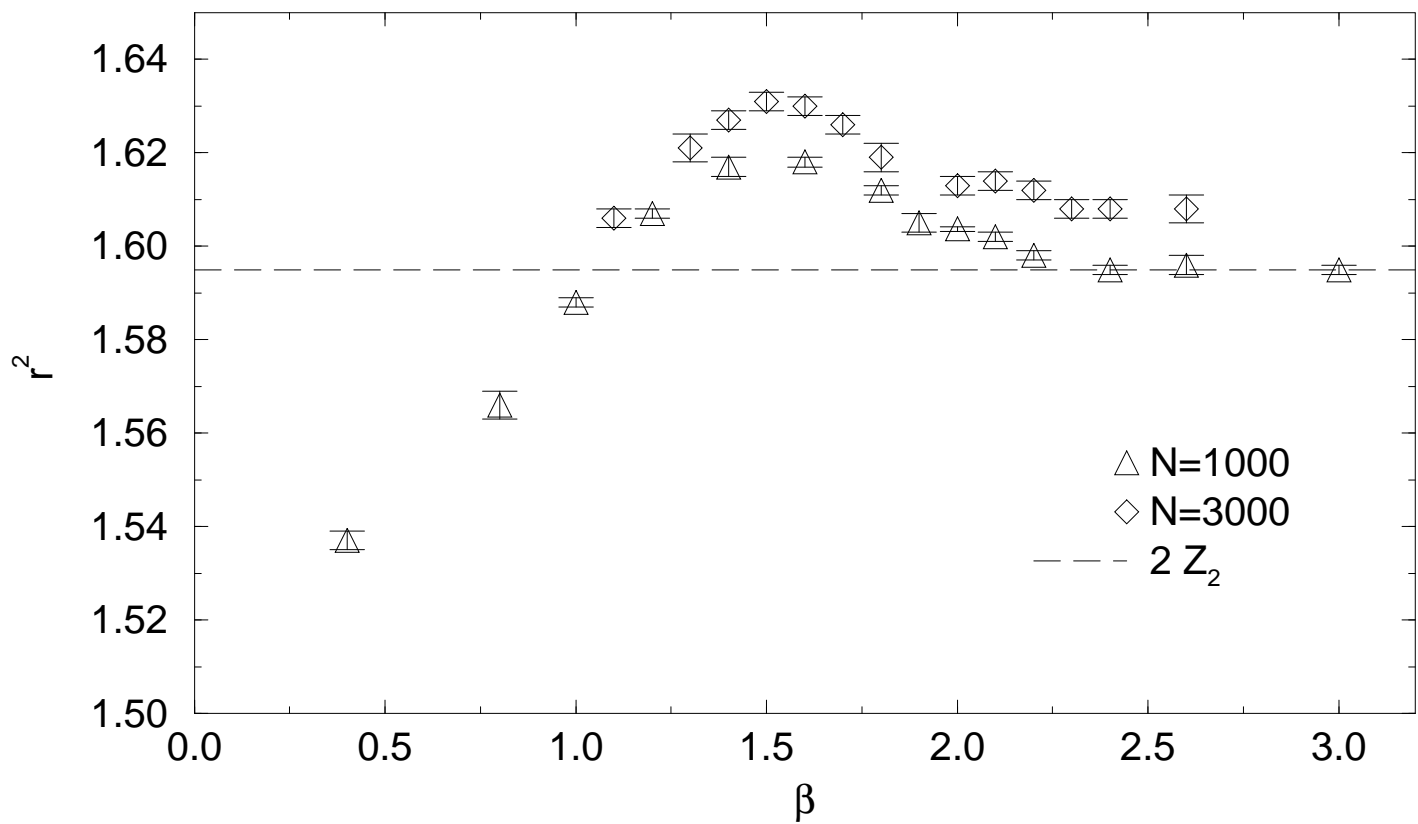
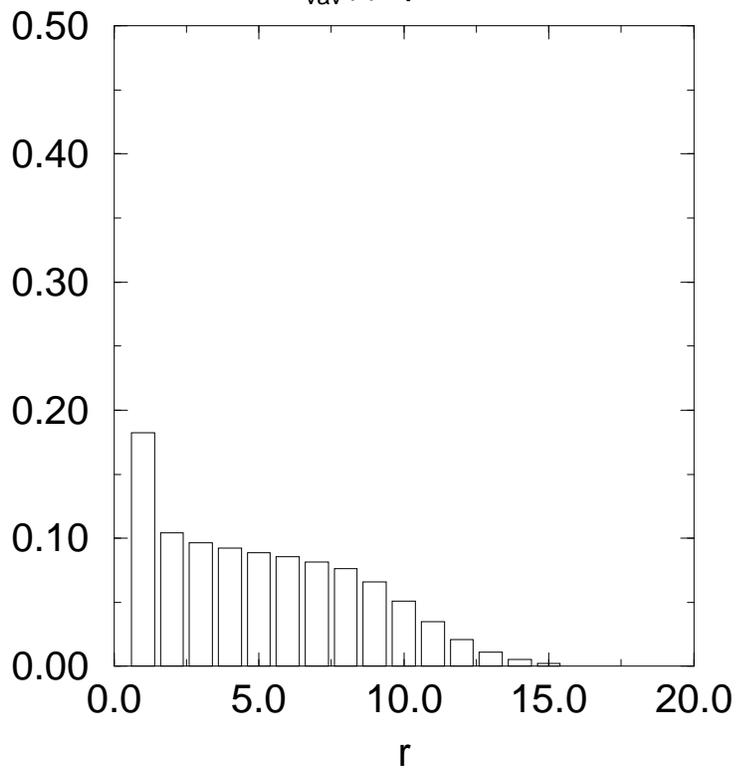
